\documentclass[amsmath,amssymb,twocolumn,
		    superscriptaddress]{revtex4}

\usepackage{color,graphicx}

\newcommand{\ket}[1]{\vert#1\rangle}
\newcommand{\bra}[1]{\langle#1\vert}
\newcommand{\CC}{{\mathbb{C}}}
\newcommand{\PostPsiP}{\textsf{Post$\mathsf{\Psi}$P}}

\newcommand{\PsiP}{\textsf{$\mathsf{\Psi}$P}}
\newcommand{\PP}{\textsf{PP}}
\newcommand{\BQP}{\textsf{BQP}}
\newcommand{\QMA}{\textsf{QMA}}
\newcommand{\NP}{\textsf{NP}}
\newcommand{\sharpP}{\textsf{\#P}}
\newcommand{\PostBQP}{\textsf{PostBQP}}

\newcommand{\poly}{\mathrm{poly}}

\begin{document}

\title{The computational complexity of PEPS}

\author{Norbert Schuch}
\affiliation{Max-Planck-Institut f\"ur Quantenoptik,
  Hans-Kopfermann-Str.\ 1, D-85748 Garching, Germany.}
\author{Michael M.\ Wolf}
\affiliation{Max-Planck-Institut f\"ur Quantenoptik,
  Hans-Kopfermann-Str.\ 1, D-85748 Garching, Germany.}
\author{Frank Verstraete}
\affiliation{Fakult\"at f\"ur Physik, Universit\"at Wien, 
Boltzmanngasse 5, A-1090 Wien, Austria.} 
\author{J.\ Ignacio Cirac}
\affiliation{Max-Planck-Institut f\"ur Quantenoptik,
  Hans-Kopfermann-Str.\ 1, D-85748 Garching, Germany.}

\begin{abstract}
 We determine the computational power of preparing
Projected Entangled Pair States (PEPS), as well as the complexity of
classically simulating them, and generally the complexity of contracting
tensor networks. While creating PEPS allows to solve \PP\  problems,
the latter two tasks are both proven to be \sharpP-complete. We
further show how PEPS can be used to approximate ground states of gapped
Hamiltonians, and that creating them is easier than creating arbitrary
PEPS.  
The main tool for our proofs is a duality between PEPS and
postselection which allows to use existing results from quantum compexity.
\end{abstract}

\maketitle

\section{Introduction}%
Computing the properties of correlated quantum many-body systems is a
central task in many fields in physics. Its complexity stems mainly from
the large dimension of the Hilbert space which grows exponentially in the
system size. In the last decades, the Density Matrix Renormalization Group
(DMRG) method has proven extremely successful in the description of
one-dimensional phenomena~\cite{white:DMRG-PRL}. Recently, it has been
shown that from the perspective of quantum information, DMRG can be
described as a variational method over the class of Matrix Product States
(MPS)~\cite{schollwoeck:rmp}. MPS structure the state space into
a hierarchy of states with polynomial description
complexity~\cite{david:mps-representations}, and it turns out that already
the lowest levels of this hierarchy approximate many physical states of
interest extremely well. MPS have a natural extension to two and higher
dimensional lattices, called Projected Entangled Pair States (PEPS), which
also have an efficient description and are promising candidates for
variational methods in higher dimensions~\cite{frank:2D-dmrg}.  It has
been shown that MPS can be created efficiently by a quantum
computer~\cite{schoen:hen-and-egg}, and that they also can be simulated
efficiently classically~\cite{vidal:simulation-of-comput}.  In contrast,
in two or more dimensions it seems to be hard to create arbitrary PEPS, as
well as to classically compute expectation values. In fact, it has 
been shown that there exist 2D PEPS which encode solutions to NP-complete
problems~\cite{frank:comp-power-of-peps}, thus posing lower bounds on
their complexity and computational power.

In the present work, we determine both the power of creating PEPS and the
complexity of classically simulating them.  We investigate which kind of
problems we could solve if we had a way to efficiently create PEPS, and 
find that these are exactly the problems in the complexity class \PP\
(deciding whether a boolean formula has more satisfying than
non-satisfying assingments). Second, we show that classically computing
local expectation values on PEPS is a \sharpP-complete problem (counting
the satisfying assignments of a boolean formula). This result can be
extended to the contraction of arbitrary tensor networks, which turns out
to be \sharpP-comlete as well.

The main tool in our proofs is a duality between PEPS and postselection,
which permits to use existing results from quantum
complexity~\cite{aaronson:postsel}: any PEPS can
be created by a postselected quantum circuit, and any output of such a
circuit can be written as a PEPS. We also apply this duality to show
that ground states of gapped local Hamiltonians in $D$ dimensions can be
efficiently approximated by the boundary of a $D+1$-dimensional PEPS. 
Finally, we compare the power of creating PEPS to the power of creating
ground states of local Hamiltonians. While in general they are equally
hard, we find that when restricting to gapped Hamiltonians, creating
ground states becomes easier: it is in the weaker class \QMA, the quantum
analogue of \NP.

\section{PEPS and postselection}%
We start by recalling the definition of PEPS~\cite{frank:mbc-peps}. 
  Consider an arbitrary undirected graph where each of the
vertices corresponds to a quantum system (a \emph{spin}) of Hilbert space
dimension $d$. A PEPS on these $N$ spins is constructed by placing as
many virtual spins of dimension $D$ on each vertex as there are adjacent
edges. Along each edge, these virtual spins form maximally entanged states
$\sum_{i=1}^D\ket i\ket i$.  The physical spins are now obtained from the
virtual ones by applying a linear map
$P^{[v]}:\CC^D\otimes\ldots\otimes\CC^D\rightarrow\CC^d$ at each vertex
$v$.  For the sake of readability, we will mostly supress the dependence
of $P$ on $v$. The graph underlying the PEPS will usually be chosen
according to the physical setup, typically a two or higher
dimensional lattice.

Let us now turn to \emph{postselected quantum
circuits}~\cite{aaronson:postsel}.  Roughly
speaking, postselection means that we can measure a qubit with the promise
of obtaining a certain outcome. More precisely, the postselected circuits
we consider start from the $\ket{0\cdots0}$ state, perform a sequence
of unitary one- and two-qubit gates, and postselect on the first qubit
being $\ket0$. Thereby, the state
$\alpha\ket0\ket{\phi_0}+\beta\ket1\ket{\phi_1}$ is projected onto the
state $\ket{\phi_0}$, which is the state created by the postselected
quantum circuit~\footnote{
We do not impose polynomial-size and uniformity conditions on the circuit,
which would yield a natural extension \PostPsiP\ of the class \PsiP\
defined in~\cite{aaronson:state-classes}.  The reason is that we will show
that there exists a uniform and efficient \emph{transformation} between
PEPS and postselected quantum states, although none of the two has to
satisfy any efficiency or uniformity condition.  
}.
Note that a state with $\alpha=0$ will not be considered a valid input.

In the following, we show that the output of a postselected quantum
circuit can be expressed efficiently as a PEPS on a 2D square lattice with
both $D=d=2$. We start by briefly recalling the concept of measurement
based quantum
computation~\cite{raussendorf:cluster-short,raussendorf:cluster-long}: One
starts from the 2D cluster state (which is a PEPS with
$D=d=2$~\cite{frank:mbc-peps}) and implements the quantum circuit by a
sequence of projective measurements on the individual spins.  Finally, the
output is found in the unmeasured qubits, up to Pauli corrections which
depend on the previous measurement outcomes. In order to express the
output of a postselected circuit as a PEPS, we therefore start by
implementing its unitary part in the measurement based
model.  We do this by projecting each
qubit on the outcome $\ket{a}$ which does \emph{not} give a Pauli
correction, by replacing the original cluster projector $P_C$ with
$\ket a\bra{a}P_C$.
This leaves us with a set of qubits holding the output of the
circuit, and by projecting the first qubit on $\ket0$,
we obtain the output of the postselected quantum circuit.  The
transformation between the representations can be carried out efficiently,
and the resulting PEPS has a size polynomial in the length of the circuit.

Conversely, any PEPS can be efficiently created by a postselected quantum
computer.  This holds for PEPS on an arbitary graph with degree (the
maximum number of edges adjacent to a vertex) at most logarithmic in the
system size, which ensures that the $P$'s are polynomial-size matrices.
The key point is that any linear map $P$ can be implemented
deterministically using postselection. To this end, append rows or colums
of zeros to make $P$ a square matrix $\tilde P$.   By appropriate
normalization, we can assume w.l.o.g.\ that $P^\dagger P\le\openone$.
Hence, there exists a unitary $U$ on the original system and one ancilla
such that $\bra 0_{\mathrm{anc}}U\ket{0}_{\mathrm{anc}}=\tilde P$. This
is, by adding an ancilla $\ket0_\mathrm{anc}$, performing $U$ and
postselecting the ancilla we can implement $\tilde P$.  In order to
generate a PEPS using postselection, we thus have to encode each of the
virtual spins in $\lceil\log D\rceil$ qubits, create the maximally
entangled pairs, and implement the $U$'s corresponding to the maps $P$,
which can be all done efficiently. We are thus left with $N$ ancillas, all
of which we have to postselect on $\ket0$. This, however, can be done with
a single postselection by computing the \textsc{or} of all ancillas into a
new ancilla and postselecting it on $\ket0$.

In summary, on the one side we have that any postselected quantum circuit
can be translated efficiently into a 2D PEPS with $D=d=2$, while
conversely there is also an efficient transform from any PEPS to a
postselected quantum circuit. In turn, this shows that all the features
and the full complexity of PEPS can already be found in the simplest case
of two-dimensional PEPS, making them an even more interesting subject for
investigations.

\section{The power of creating PEPS}%
Let us first briefly
introduce the complexity classes \sharpP\ and
\PP~\cite{papadimitriou:book}.  Consider an efficiently computable boolean
function $f:\{0,1\}^N\rightarrow\{0,1\}$, and let $s\equiv
s(f):=|\{x:f(x)=1\}|$ be the number of satisfying assignments.  Then,
finding $s$ defines the counting class \sharpP, while determining whether
$s\ge 2^{n-1}$ (i.e., finding the first bit of $s$) defines the decision
class \PP.  This class contains \NP\ and \BQP\ as well as \QMA, the
quantum version of \NP.

\begin{figure}[b]
\includegraphics[width=.90\columnwidth]{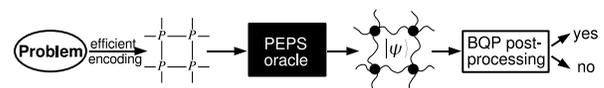}
\caption{\label{fig:power-of-peps}
The power of creating PEPS: The original decision problem is transformed
into  a PEPS description by a polynomial-time algorithm. The black box
creates the corresponding quantum state, and an efficient quantum
postprocessing returns the solution. Which kind of problems can we solve
this way?}
\end{figure}
First, we investigate the computational power of creating PEPS.
More precisely, we consider the scenario of Fig.~\ref{fig:power-of-peps}:
We want to know which decision problems we can solve with one use of a
PEPS oracle, i.e., a black box which creates the quantum state from its
classical PEPS description,  together with efficient classical
pre-processing and quantum post-processing.

We now use the PEPS--postselection duality to show that the power of
creating PEPS equals \PP.  It has been shown that \textsf{PostBQP}---the
class of decision problems which can be solved by a postselected quantum
computer---equals \PP, $\mbox\PostBQP=\mbox\PP$~\cite{aaronson:postsel}.
This readily implies that a PEPS oracle allows us to solve \PP\ problems
instantaneously by preparing the output of the postselected circuit as a
PEPS and just measuring one output qubit in the computational basis. On
the other hand, this is the best we can do with a single use of the PEPS
oracle,  since every PEPS can be generated efficiently by a postselected
quantum computer. \BQP\ postprocessing instead of a simple one-qubit
measurement does not increase the computational power, since it commutes
with the postselection and can thus be incorporated in the PEPS.

The fact that creating PEPS allows to solve \PP-complete problems strongly
suggests the existence of PEPS which cannot be created efficiently by a
quantum computer.  Note however that the states which appear in the
\PP-hardness proof above are not of this type: once the corresponding
counting problem is solved, they can be easily constructed. While it
appears very unlikely that all PEPS can be constructed efficiently from
some normal form (it would imply
$\mbox{\textsf{QMA}}=\mbox{\textsf{QCMA}}$ and
$\mbox{\textsf{BQP/qpoly}}=\mbox{\textsf{BQP/poly}}$~%
\cite{aaronson:qu-vs-class-advice}), an example of such a state is still
missing.

\section{The classical complexity of PEPS}%
Let us now investigate the complexity of classically simulating PEPS, and
its generalization, the contraction of  tensor networks.
For the case of PEPS, there are at least three possible definitions of the
problem: compute the normalization of the PEPS (\textsc{norm}), compute
the unnormalized expectation value of some observable (\textsc{uev}), and
compute the normalized expectation value (\textsc{nev}).  Since they can
be transformed easily into each other~\footnote{
\label{footnote:reductions}
\textsc{uev} of $\openone$ gives \textsc{norm}, while \textsc{uev} of an
operator $A$ is obtained by applying \textsc{norm} twice, $\bra\psi A
\ket\psi=\langle\tilde\psi\ket{\tilde\psi}-\|A\|\bra\psi\psi\rangle$. Here,
$\ket{\tilde\psi}$ is derived from $\ket{\psi}$ by replacing the relevant
$P$ by $(A+\|A\|\openone)^{1/2}P$.  Clearly, \textsc{uev} and
\textsc{norm} allow to compute \textsc{nev}. Conversely, to compute the
norm of a PEPS write it as a quantum circuit, but stop before the
postselection. Then, its norm equals the \textsc{nev} of
$\mathrm{diag(1,0)}$ on the qubit to be postselected, which equals the
\textsc{nev} on a PEPS.  All reductions are weakly parsimonious: problem A
can be solved by \emph{one} call to problem B, with efficient
pre-processing of the input \emph{and} post-processing of the output.
Note that two (or more) parallel \sharpP-queries can be encoded in a
single one, by considering $h(x,y,b)$, defined as $f(x)$ for $b=0$ and
$g(y)$ for $b=1$, $x=0$.  
}, 
we will use whichever is most appropriate.

We first show that contracting PEPS is \sharpP-hard, i.e., that for any
(polynomial) boolean function $f$, $s(f)$ can be found by
simulating a PEPS. Therefore, we take a quantum circuit which creates
$\sum_x\ket{x}_A\ket{f(x)}_B$ and encode it in a PEPS.  Then, the
normalized expectation value of $\sigma_z$ of $B$ allows to compute
$s(f)$. 

To show that the simulation of PEPS is inside \sharpP, we have to show
that the normalization of the PEPS, or equivalently the success
probability for the postselection, can be computed by counting the satisfying
assigments of some boolean function. This can be done by adapting
well-established quantum complexity techniques (see \cite{aaronson:postsel} and
references therein):  First, approximate the postselected circuit using
only Toffoli and Hadamard
gates~\cite{shi:toffoli-hadamard,aharonov:toffoli-hadamard}.  The
probability $p_x$ for a state $\ket{x}$ before postselection is obtained
as a kind of path integral~\cite{dawson:pathint},  by 
summing the amplitudes for all possible ``computational paths''
$\zeta=(\zeta_1,\ldots,\zeta_{T-1})$, where $\ket{\zeta_t}$ is the state
step $t$ and $T$ the length of the circuit:
$$ p_x=|\sum_\zeta\alpha_{x,\zeta}|^2=\sum_{\zeta,\zeta'}
\alpha_{x,\zeta}\alpha^*_{x,\zeta'}\ , $$ with $\alpha_{x,\zeta}$ 
a product over transition amplitudes $A_{\zeta_{t}\rightarrow\zeta_{t+1}}$
along the path $\zeta$.  The normalization of the PEPS is obtained
as the sum over all states where the postselection succeeds, $\sum_{\bar
x}p_{(0,\bar x)}$.  This can be rewritten as the sum over an efficiently
computable function
$f(\bar x,\zeta,\zeta')= \alpha_{(0,\bar x),\zeta}\alpha^*_{(0,\bar
x),\zeta'}$ which takes values in $\{0,\pm1\}$, as  the circuit consisted
only of Toffoli and Hadamard gates. Now this sum
can be computed by counting the satisfying assigments of the function
$f_\mathrm{bool}(\xi,z):=(f(\xi)\ge z)$, $z\in\{0,1\}$, which shows that
the simulation of PEPS is in \sharpP.  Together, we find that the
classical simulation of PEPS is \sharpP-complete under weakly parsimonious
reductions (see~[23]).

It is natural to ask whether this also shows that contracting general
tensor networks is in \sharpP. For a tensor network $T$, let us denote its
contraction by $\mathcal C(T)\in\mathbb C$. Since the contraction of PEPS
is a special case, it is clear that the problem is \sharpP-hard. To place
it within \sharpP, observe first that $|\mathcal C(T)|^2=\mathcal
C(T\otimes T^*)$ can be found by attaching a physical system of dimension
one to each site and computing the normalization of the resulting PEPS.
To determine the phase of $\mathcal C(T)$, observe that $\mathcal
C(T\oplus T')=\mathcal C(T)+\mathcal C(T')$.  Thus, by setting $T'=T^*$,
we get $|\mathrm{Re}(\mathcal C(T))|$, while the sign can be determined by
adding another $T''\equiv c>0$. This proves  that contracting tensor
networks is \sharpP-complete.

The obtained hardness results are stable under approximations. To see why,
note that any counting problem can be reduced to any of our three
primitives with only linear postprocessing, and thus approximating these
primitives is as hard as approximating counting problems can be. For
\textsc{nev}, this again works by preparing $\sum \ket x_A \ket{f(x)}_B$
and computing the expectation value of $B$. For \textsc{norm} and thus
\textsc{uev}, note that the output of any normal quantum circuit and thus
$\sum \ket x_A\ket{f(x)}_B$ has a known norm when written as a PEPS, since
the success probability of each cluster projector is known, and the
probability of the two measurement outcomes in the cluster is
unbiased~\cite{raussendorf:cluster-long}.  Thus, the probability for $\ket
1_B$ can be readily determined from the norm of the PEPS where we
postselected on $\ket 1_B$.

\section{PEPS and ground states}%
The interest in MPS and PEPS stems mainly from the fact that those states
perform extremely well in approximating ground states. In the following,
we use the PEPS--postselection duality, and a relation between
postselection and cooling, to shed new light on the connection between
PEPS and ground states. In particular, we show that the unique ground
state of a gapped Hamiltonian on a $D$-dimensional lattice can be
approximated efficiently by the border of a PEPS with $D+1$ dimensions.

Consider a Hamiltonian on $N$ spins, $H=\sum_i H_i$, where each $H_i$ acts
on a finite number of spins, with a unique ground state and a polynomial
energy gap $\Delta\ge1/\mathrm{poly}(N)$. Starting from a random state
$\ket\chi$, the ground state can be efficiently approximated via
$\ket{\psi_0}\approx\exp[-\beta H]\ket\chi$.  The imaginary time evolution
can in turn be approximated using the Trotter decomposition, which only
requires operations $\exp[-\beta/N H_i]$ acting on finitely many spins.
Since those operations are linear, they can be implemented using
postselection, and we see that postselection can be used to cool into the
ground state.  By embedding the postselected cooling procedure in a PEPS,
the ground state of any gapped $N$-particle Hamiltonian can be
approximated up to $\epsilon$ by the boundary of a PEPS, where the extra
dimension has depth $M\sim\poly(N,1/\epsilon)$~\footnote{ One might object
that the performance of 1D variational methods is much better. However,
there are several differences: Our method works for any dimension, it is
constructive, it does not break translational symmetry, and it implements
the complete evolution $\exp[-\beta H]$.  }.  In case the $H_i$ are local,
the PEPS can be simplified considerably since any local linear operation
can be implemented directly on the level of the PEPS without the need for
ancilla qubits.

\section{The power of creating ground states}%
As we have seen, PEPS can encapsulate problems as hard as \PP. However,
these PEPS are quite artificial, while in practice one is often interested
in PEPS in connection with ground states.  Therefore, let us have a look
at the computational power of a ground state oracle, i.e., a black box
which creates the ground state from the Hamiltonian.

First, let us introduce the complexity class \QMA~\cite{kitaev:qma}.
Colloquially, \QMA\ is the quantum version of \NP, i.e., it contains all
decision problems where for the ``yes'' instance, there exists an
efficiently checkable \emph{quantum} proof, while there is no proof for
any ``no'' instance. In a seminal work,
Kitaev~\cite{kitaev:book,kitaev:qma} has shown
that the problem of determining ground state energies of local
Hamiltonians up to polynomial accuracy is \QMA-complete. More precisely,
in \textsc{local hamiltonian} one is given an $N$-qubit local Hamiltonian
$H=\sum H_i$ with the promise that the ground state energy $E_0<a$  or
$E_0>b$, $b-a>1/\mathrm{poly}(N)$, and the task is to decide whether
$E_0<a$.  Clearly, the ground state of $H$ serves as a proof for a ``yes''
instance.  In successive works, the class of Hamiltonians has been
restricted down to two-particle neareast neighbor Hamiltonians on a 2D
lattice of qubits~\cite{terhal:lh-2d-qma}.

Let us briefly reconsider our cooling protocol in the light of \QMA. It is
easy to see that the \QMA\ proof need not necessarily be the ground state,
as long as it is close enough in energy (depending on the verifier).  Since
our cooling protocol suppresses higher energy levels
exponentially, the correspondence between postselection and cooling shows
that a postselected quantum computer can be used to create proofs for
\QMA\ problems, or differently speaking, that any \QMA\ proof can be
efficiently expressed as a PEPS.

In the following, we give some observations which indicate that creating
ground states of gapped Hamiltonians is easier than creating PEPS.  First,
note that a ground state oracle for arbitrary Hamiltonians is still as
powerful as \PP. To see why, take a \PP\ problem and encapsulate it in a
PEPS. By perturbing the $P$'s randomly by a small amount, one obtains a
PEPS which is the unique ground state of a local Hamiltonian, which can
be derived efficiently from the
$P$'s~\cite{david:mps-representations,david:mps-rep-2d}.
This shows that an unrestricted ground state oracle enables us to solve
\PP\ problems.  However, the gap $\Delta$ of the above Hamiltonian will be
exponentially small: if not, one could add a small penalty, say
$\Delta/100$, on the ``answer'' qubit, and use that the original
Hamiltonian has ground state energy $E_0=0$: Then, determining the value
of that qubit could be solved in \QMA, thus proving $\mbox\QMA=\mbox\PP$
which is considered unlikely~\cite{vyalyi:qma-pp}.

Since ground states of general Hamiltonians are not easier to create than
PEPS, let us now assume an oracle which only works for local Hamiltonians
with a unique ground state, known ground state energy, and a polynomial
spectral gap to the first excited state.  (Alternatively, one could
consider ``proof oracles'' for the \textsc{local hamiltonian} problem.) It
is easy to see that this restricted oracle, even with \BQP\
postprocessing, is at most as powerful as \QMA. The proof is the
ground state, and the verifier is constructed as follows. Let $V_1$ be the
verifier for the ground state, it accepts the ground state with
$p_\mathrm{GS}$, and any excited state with probability at most
$p_\mathrm{ES}=p_\mathrm{GS}-\Delta$, $\Delta=1/\poly(N)$. Further, let
$V_2$ be the postprocessing circuit which has a polynomial separation
between the ``yes'' and the ``no'' answer if applied to the ground state, 
$p_\mathrm{yes}=1/2+\delta$,
$\delta=1/\poly(N)$. Take $Q=\frac{\Delta/2+1}{\Delta+1}$, and construct the
complete verifier as follows: with probability $Q$, run $V_1$, and with
$(1-Q)$, run $V_2$. One can readily check that this gives a polynomial
separation between the cases where the proof is the ground state
\emph{and} the postprocessing return ``yes'', and the cases where either
the proof is not the ground state or the postprocessing returns ``no''.
The same strategy can be used to show that a PEPS oracle cannot be tested
on all inputs unless $\mbox\QMA=\mbox\PP$: Otherwise, one could take a
\PP-hard PEPS and construct a verifier which either runs the testing
routine or reads out the \PP\ solution.

These observations show that imposing a constraint on the spectral gap of
a Hamiltonian has direct implications on its computational complexity, and
we think that the complexity properties of gapped Hamiltonians are worth
being considered. On the one side, in the above scenario it is not clear
whether all \QMA\ problems can be solved using this oracle, on the other
side, it is not clear how important the knowledge of the ground state
energy is---note that we however also had this knowledge in the \PP-hard
case. It is also an interesting question whether the problem \textsc{local
hamiltonian} remains \QMA-complete when restricting to polynomially gapped
Hamiltonians. If not, \textsc{gapped local hamiltonian} should be a
natural candidate for a physically motivated class of problems weaker than
\QMA.

\section{Acknowledgements}%
We thank C.\ Dawson, J.\ Eisert, D.\ P{\'e}rez Garc{\'\i}a, T.\ Roscilde,
K.~G.\ Vollbrecht, and one of the referees for helpful discussions and
comments. This work was supported by the Elite Network of Bavaria (ENB)
project QCCC, and by the DFG-Forschergruppe 635.


\begin{thebibliography}{22}
\expandafter\ifx\csname natexlab\endcsname\relax\def\natexlab#1{#1}\fi
\expandafter\ifx\csname bibnamefont\endcsname\relax
  \def\bibnamefont#1{#1}\fi
\expandafter\ifx\csname bibfnamefont\endcsname\relax
  \def\bibfnamefont#1{#1}\fi
\expandafter\ifx\csname citenamefont\endcsname\relax
  \def\citenamefont#1{#1}\fi
\expandafter\ifx\csname url\endcsname\relax
  \def\url#1{\texttt{#1}}\fi
\expandafter\ifx\csname urlprefix\endcsname\relax\def\urlprefix{URL }\fi
\providecommand{\bibinfo}[2]{#2}
\providecommand{\eprint}[2][]{\url{#2}}

\bibitem[{\citenamefont{White}(1992)}]{white:DMRG-PRL}
\bibinfo{author}{\bibfnamefont{S.~R.} \bibnamefont{White}},
  \bibinfo{journal}{Phys. Rev. Lett.} \textbf{\bibinfo{volume}{69}},
  \bibinfo{pages}{2863} (\bibinfo{year}{1992}).

\bibitem[{\citenamefont{Schollw{\"o}ck}(2005)}]{schollwoeck:rmp}
\bibinfo{author}{\bibfnamefont{U.}~\bibnamefont{Schollw{\"o}ck}},
  \bibinfo{journal}{Rev.\ Mod.\ Phys.} \textbf{\bibinfo{volume}{77}},
  \bibinfo{pages}{259} (\bibinfo{year}{2005}), \eprint{cond-mat/0409292}.

\bibitem[{\citenamefont{Perez-Garcia et~al.}(2006)\citenamefont{Perez-Garcia,
  Verstraete, Wolf, and Cirac}}]{david:mps-representations}
\bibinfo{author}{\bibfnamefont{D.}~\bibnamefont{Perez-Garcia}},
  \bibinfo{author}{\bibfnamefont{F.}~\bibnamefont{Verstraete}},
  \bibinfo{author}{\bibfnamefont{M.~M.} \bibnamefont{Wolf}}, \bibnamefont{and}
  \bibinfo{author}{\bibfnamefont{J.~I.} \bibnamefont{Cirac}}
  (\bibinfo{year}{2006}), \eprint{quant-ph/0608197}.

\bibitem[{\citenamefont{Verstraete and
  Cirac}(2004{\natexlab{a}})}]{frank:2D-dmrg}
\bibinfo{author}{\bibfnamefont{F.}~\bibnamefont{Verstraete}} \bibnamefont{and}
  \bibinfo{author}{\bibfnamefont{J.~I.} \bibnamefont{Cirac}}
  (\bibinfo{year}{2004}{\natexlab{a}}), \eprint{cond-mat/0407066}.

\bibitem[{\citenamefont{Sch{\"o}n et~al.}(2005)\citenamefont{Sch{\"o}n, Solano,
  Verstraete, Cirac, and Wolf}}]{schoen:hen-and-egg}
\bibinfo{author}{\bibfnamefont{C.}~\bibnamefont{Sch{\"o}n}},
  \bibinfo{author}{\bibfnamefont{E.}~\bibnamefont{Solano}},
  \bibinfo{author}{\bibfnamefont{F.}~\bibnamefont{Verstraete}},
  \bibinfo{author}{\bibfnamefont{J.~I.} \bibnamefont{Cirac}}, \bibnamefont{and}
  \bibinfo{author}{\bibfnamefont{M.~M.} \bibnamefont{Wolf}},
  \bibinfo{journal}{Phys.\ Rev.\ Lett.} \textbf{\bibinfo{volume}{95}},
  \bibinfo{pages}{110503} (\bibinfo{year}{2005}), \eprint{quant-ph/0501096}.

\bibitem[{\citenamefont{Vidal}(2003)}]{vidal:simulation-of-comput}
\bibinfo{author}{\bibfnamefont{G.}~\bibnamefont{Vidal}},
  \bibinfo{journal}{Phys.\ Rev.\ Lett.} \textbf{\bibinfo{volume}{91}},
  \bibinfo{pages}{147902} (\bibinfo{year}{2003}), \eprint{quant-ph/0301063}.

\bibitem[{\citenamefont{Verstraete et~al.}(2006)\citenamefont{Verstraete, Wolf,
  Perez-Garcia, and Cirac}}]{frank:comp-power-of-peps}
\bibinfo{author}{\bibfnamefont{F.}~\bibnamefont{Verstraete}},
  \bibinfo{author}{\bibfnamefont{M.~M.} \bibnamefont{Wolf}},
  \bibinfo{author}{\bibfnamefont{D.}~\bibnamefont{Perez-Garcia}},
  \bibnamefont{and} \bibinfo{author}{\bibfnamefont{J.~I.} \bibnamefont{Cirac}},
  \bibinfo{journal}{Phys.\ Rev.\ Lett.} \textbf{\bibinfo{volume}{96}},
  \bibinfo{pages}{220601} (\bibinfo{year}{2006}), \eprint{quant-ph/0601075}.

\bibitem[{\citenamefont{Aaronson}(2005{\natexlab{a}})}]{aaronson:postsel}
\bibinfo{author}{\bibfnamefont{S.}~\bibnamefont{Aaronson}},
  \bibinfo{journal}{Proc.\ R.\ Soc.\ Lond.\ A} \textbf{\bibinfo{volume}{461}},
  \bibinfo{pages}{3473} (\bibinfo{year}{2005}{\natexlab{a}}),
  \eprint{quant-ph/0412187}.

\bibitem[{\citenamefont{Verstraete and
  Cirac}(2004{\natexlab{b}})}]{frank:mbc-peps}
\bibinfo{author}{\bibfnamefont{F.}~\bibnamefont{Verstraete}} \bibnamefont{and}
  \bibinfo{author}{\bibfnamefont{J.~I.} \bibnamefont{Cirac}},
  \bibinfo{journal}{Phys.~Rev.~A} \textbf{\bibinfo{volume}{70}},
  \bibinfo{pages}{060302} (\bibinfo{year}{2004}{\natexlab{b}}),
  \eprint{quant-ph/0311130}.

\bibitem[{\citenamefont{Raussendorf and
  Briegel}(2001)}]{raussendorf:cluster-short}
\bibinfo{author}{\bibfnamefont{R.}~\bibnamefont{Raussendorf}} \bibnamefont{and}
  \bibinfo{author}{\bibfnamefont{H.~J.} \bibnamefont{Briegel}},
  \bibinfo{journal}{Phys. Rev. Lett.} \textbf{\bibinfo{volume}{86}},
  \bibinfo{pages}{5188} (\bibinfo{year}{2001}), \eprint{quant-ph/0010033}.

\bibitem[{\citenamefont{Raussendorf et~al.}(2003)\citenamefont{Raussendorf,
  Browne, and Briegel}}]{raussendorf:cluster-long}
\bibinfo{author}{\bibfnamefont{R.}~\bibnamefont{Raussendorf}},
  \bibinfo{author}{\bibfnamefont{D.~E.} \bibnamefont{Browne}},
  \bibnamefont{and} \bibinfo{author}{\bibfnamefont{H.~J.}
  \bibnamefont{Briegel}}, \bibinfo{journal}{Phys.~Rev.~A}
  \textbf{\bibinfo{volume}{68}}, \bibinfo{pages}{022312}
  (\bibinfo{year}{2003}), \eprint{quant-ph/0301052}.

\bibitem[{\citenamefont{Papadimitriou}(1994)}]{papadimitriou:book}
\bibinfo{author}{\bibfnamefont{C.~M.} \bibnamefont{Papadimitriou}},
  \emph{\bibinfo{title}{{Computational complexity}}}
  (\bibinfo{publisher}{Addison-Wesley}, \bibinfo{address}{Reading, MA},
  \bibinfo{year}{1994}).

\bibitem[{\citenamefont{Aaronson}(2005{\natexlab{b}})}]{aaronson:qu-vs-class-a%
dvice}
\bibinfo{author}{\bibfnamefont{S.}~\bibnamefont{Aaronson}},
  \bibinfo{journal}{Theory of Computing} \textbf{\bibinfo{volume}{1}},
  \bibinfo{pages}{1} (\bibinfo{year}{2005}{\natexlab{b}}),
  \eprint{quant-ph/0402095}.

\bibitem[{\citenamefont{Shi}(2003)}]{shi:toffoli-hadamard}
\bibinfo{author}{\bibfnamefont{Y.}~\bibnamefont{Shi}},
  \bibinfo{journal}{Quant.\ Inf.\ Comput.} \textbf{\bibinfo{volume}{3}},
  \bibinfo{pages}{84} (\bibinfo{year}{2003}), \eprint{quant-ph/0205115}.

\bibitem[{\citenamefont{Aharonov}(2003)}]{aharonov:toffoli-hadamard}
\bibinfo{author}{\bibfnamefont{D.}~\bibnamefont{Aharonov}}
  (\bibinfo{year}{2003}), \eprint{quant-ph/0301040}.

\bibitem[{\citenamefont{Dawson et~al.}(2005)\citenamefont{Dawson, Haselgrove,
  Hines, Mortimer, Nielsen, and Osborne}}]{dawson:pathint}
\bibinfo{author}{\bibfnamefont{C.~M.} \bibnamefont{Dawson}},
  \bibinfo{author}{\bibfnamefont{H.~L.} \bibnamefont{Haselgrove}},
  \bibinfo{author}{\bibfnamefont{A.~P.} \bibnamefont{Hines}},
  \bibinfo{author}{\bibfnamefont{D.}~\bibnamefont{Mortimer}},
  \bibinfo{author}{\bibfnamefont{M.~A.} \bibnamefont{Nielsen}},
  \bibnamefont{and} \bibinfo{author}{\bibfnamefont{T.~J.}
  \bibnamefont{Osborne}}, \bibinfo{journal}{Quant.\ Inf.\ Comput.}
  \textbf{\bibinfo{volume}{5}}, \bibinfo{pages}{102} (\bibinfo{year}{2005}),
  \eprint{quant-ph/0408129}.

\bibitem[{\citenamefont{Aharonov and Naveh}(2002)}]{kitaev:qma}
\bibinfo{author}{\bibfnamefont{D.}~\bibnamefont{Aharonov}} \bibnamefont{and}
  \bibinfo{author}{\bibfnamefont{T.}~\bibnamefont{Naveh}}
  (\bibinfo{year}{2002}), \eprint{quant-ph/0210077}.

\bibitem[{\citenamefont{Kitaev et~al.}(2002)\citenamefont{Kitaev, Shen, and
  Vyalyi}}]{kitaev:book}
\bibinfo{author}{\bibfnamefont{A.~Y.} \bibnamefont{Kitaev}},
  \bibinfo{author}{\bibfnamefont{A.~H.} \bibnamefont{Shen}}, \bibnamefont{and}
  \bibinfo{author}{\bibfnamefont{M.~N.} \bibnamefont{Vyalyi}},
  \emph{\bibinfo{title}{Classical and quantum computation}}
  (\bibinfo{publisher}{American Mathematical Society},
  \bibinfo{address}{Providence, Rhode Island}, \bibinfo{year}{2002}).

\bibitem[{\citenamefont{Oliveira and Terhal}(2005)}]{terhal:lh-2d-qma}
\bibinfo{author}{\bibfnamefont{R.}~\bibnamefont{Oliveira}} \bibnamefont{and}
  \bibinfo{author}{\bibfnamefont{B.~M.} \bibnamefont{Terhal}}
  (\bibinfo{year}{2005}), \eprint{quant-ph/0504050}.

\bibitem[{\citenamefont{Perez-Garcia et~al.}()\citenamefont{Perez-Garcia, Wolf,
  Verstraete, and Cirac}}]{david:mps-rep-2d}
\bibinfo{author}{\bibfnamefont{D.}~\bibnamefont{Perez-Garcia}},
  \bibinfo{author}{\bibfnamefont{M.~M.} \bibnamefont{Wolf}},
  \bibinfo{author}{\bibfnamefont{F.}~\bibnamefont{Verstraete}},
  \bibnamefont{and} \bibinfo{author}{\bibfnamefont{J.~I.} \bibnamefont{Cirac}},
  \emph{\bibinfo{title}{in prepraration}}.

\bibitem[{\citenamefont{Vyalyi}(2003)}]{vyalyi:qma-pp}
\bibinfo{author}{\bibfnamefont{M.}~\bibnamefont{Vyalyi}},
  \bibinfo{journal}{Electronic Colloquium on Computational Complexity}
  \textbf{\bibinfo{volume}{10}} (\bibinfo{year}{2003}),
  \url{http://eccc.hpi-web.de/}
  \url{eccc-reports/2003/TR03-021/index.html}.

\bibitem[{\citenamefont{Aaronson}(2004)}]{aaronson:state-classes}
\bibinfo{author}{\bibfnamefont{S.}~\bibnamefont{Aaronson}},
  \bibinfo{journal}{Proc.\ ACM STOC} pp. \bibinfo{pages}{118--127}
  (\bibinfo{year}{2004}), \eprint{quant-ph/0311039}.

\end{thebibliography}
\end{document}